\documentclass[10pt,conference]{IEEEtran}

\usepackage{graphicx,epsfig,subfigure}
\usepackage{times}
\usepackage{amsmath,amssymb,latexsym}
\usepackage{setspace}
\usepackage{longtable}
\usepackage{multicol}
\usepackage{cite}
\usepackage{color}

\newtheorem{theorem}{Theorem}

\newtheorem{remark}{Remark}

 \providecommand{\Bv}{\mathbf{B}}

 \providecommand{\Fv}{\mathbf{F}}

 \providecommand{\Hv}{\mathbf{H}}
 \providecommand{\Iv}{\mathbf{I}}

 \providecommand{\Nv}{\mathbf{N}}
 \providecommand{\Mv}{\mathbf{M}}

 \providecommand{\Sv}{\mathbf{S}}
 
\providecommand{\uv}{\mathbf{u}} \providecommand{\Uv}{\mathbf{U}}
 \providecommand{\Vv}{\mathbf{V}}

\providecommand{\xv}{\mathbf{x}} \providecommand{\Xv}{\mathbf{X}}
\providecommand{\yv}{\mathbf{y}} \providecommand{\Yv}{\mathbf{Y}}
 \providecommand{\Zv}{\mathbf{Z}}

 \providecommand{\Cc}{{\mathcal C}}

 \providecommand{\Rc}{{\mathcal R}}

\providecommand{\Yt}{\widetilde{\mathbf{Y}}}
\providecommand{\yt}{\widetilde{\mathbf{y}}}
\providecommand{\Zt}{\widetilde{\mathbf{Z}}}

\providecommand{\Nt}{\widetilde{\mathbf{N}}}

\providecommand{\T}{\intercal}

\begin{document}
\title{MIMO Gaussian Broadcast Channels with Confidential and Common Messages}
\IEEEoverridecommandlockouts

\author{Ruoheng Liu, Tie Liu, H. Vincent Poor, and Shlomo Shamai (Shitz)%
\thanks{This research was supported by the United States National Science Foundation under Grant
CNS-09-05398, CCF-08-45848 and CCF-09-16867, by the Air Force Office
of Scientific Research under Grant FA9550-08-1-0480, by the European
Commission in the framework of the FP7 Network of Excellence in
Wireless Communications NEWCOM++,
and by the Israel Science Foundation.}%
\thanks{Ruoheng Liu and H. Vincent Poor are with the Department of Electrical Engineering,
Princeton University, Princeton, NJ 08544, USA (e-mail: \{rliu,poor\}@princeton.edu).}%
\thanks{Tie Liu is with the Department of Electrical and Computer Engineering, Texas
A\&M University, College Station, TX 77843, USA (e-mail: tieliu@tamu.edu).}%
\thanks{Shlomo Shamai (Shitz) is with the Department of Electrical Engineering,
Technion-Israel Institute of Technology, Technion City, Haifa 32000,
Israel (e-mail: sshlomo@ee.technion.ac.il).}%
}

\maketitle

\begin{abstract}
This paper considers the problem of secret communication over a
two-receiver multiple-input multiple-output (MIMO) Gaussian
broadcast channel. The transmitter has two independent, confidential
messages and a common message. Each of the confidential messages is
intended for one of the receivers but needs to be kept perfectly
secret from the other, and the common message is intended for both
receivers. It is shown that a natural scheme that combines secret
dirty-paper coding with Gaussian superposition coding achieves the
secrecy capacity region. To prove this result, a channel-enhancement
approach and an extremal entropy inequality of Weingarten {\it et
al.} are used.
\end{abstract}


\section{Introduction}\label{sec:INT}
In this paper, we study the problem of secret communication over a
multiple-input multiple-output (MIMO) Gaussian broadcast channel
with two receivers. The transmitter is equipped with $t$ transmit
antennas, and receiver~$k$, $k=1,2$, is equipped with $r_k$ receive
antennas. A discrete-time sample of the channel at time $m$ can be
written as
\begin{equation}
\Yv_{k}[m] = \Hv_k\mathbf{X}[m]+\Zv_k[m], \quad k=1,2 \label{eq:Ch1}
\end{equation}
where $\Hv_k$ is the (real) channel matrix of size $r_{k} \times t$,
and $\{\Zv_k[m]\}_m$ is an independent and identically distributed
(i.i.d.) additive vector Gaussian noise process with zero mean and
identity covariance matrix. The channel input $\{\mathbf{X}[m]\}_m$
is subject to the matrix power constraint:
\begin{equation}
\frac{1}{n}\sum_{m=1}^{n}\left(\mathbf{X}[m]\mathbf{X}^{\T}[m]\right)
\preceq \mathbf{S} \label{eq:MC}
\end{equation}
where $\mathbf{S}$ is a positive semidefinite matrix, and
``$\preceq$'' denotes ``less than or equal to'' in the positive
semidefinite partial ordering between real symmetric matrices. Note
that \eqref{eq:MC} is a rather general power constraint that
subsumes many other important power constraints including the
average total and per-antenna power constraints as special cases.

\begin{figure}[t]
 \centerline{\includegraphics[width=\linewidth,draft=false]{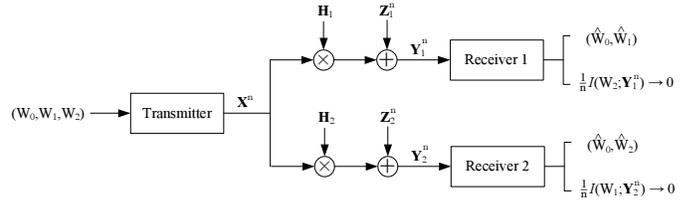}}
 \caption{Channel model}
 \label{fig:chm}
\end{figure}

As shown in Fig.~\ref{fig:chm}, we consider the communication
scenario in which there is a common message $W_0$ and two
independent, confidential messages $W_1$ and $W_2$ at the
transmitter. Message $W_0$ is intended for both receivers. Message
$W_1$ is intended for receiver 1 but needs to be kept secret from
receiver 2, and message $W_2$ is intended for receiver 1 but needs
to be kept secret from receiver 2. The confidentiality of the
messages at the unintended receivers is measured using the
normalized information-theoretic criteria \cite{CK-IT78}
\begin{align}
\frac{1}{n}I(W_1;\Yv_{2}^n) \rightarrow 0 \quad \mbox{and} \quad
\frac{1}{n}I(W_2;\Yv_{1}^n) \rightarrow 0 \label{eq:PS-1}
\end{align}
where $\Yv_k^n:=(\Yv_k[1],\ldots,\Yv_k[n])$, $k=1,2$, and the limits
are taken as the block length $n \rightarrow \infty$. The goal is to
characterize the \emph{entire} secrecy rate region $\Cc_s^{\rm
[SBC]}(\Hv_1,\Hv_2,\Sv)=\{(R_0,R_1,R_2)\}$ that can be achieved by
any coding scheme, where $R_0$, $R_1$ and $R_2$ are the
communication rates corresponding to the common message $W_0$, the
confidential message $W_1$ destined for receiver 1 and the
confidential message $W_2$ destined for receiver 2, respectively.

In recent years, MIMO secret communication has been an active area
of research. Several \emph{special} cases of the communication
problem that we consider here have been studied in the literature.
Specifically,
\begin{itemize}
\item With only one confidential message ($W_1$ or $W_2$), the problem
was studied as the MIMO Gaussian wiretap channel. The secrecy
capacity of the MIMO Gaussian wiretap channel was characterized in
\cite{LS-IT09} and \cite{BLVS-EURASIP09} under the matrix power
constraint \eqref{eq:MC} and in \cite{KW-Allerton07} and
\cite{OH-ISIT08} under an average total power constraint.
\item With both confidential messages $W_1$ and $W_2$ but \emph{without}
the common message $W_0$, the problem was studied in \cite{LP-IT09}
for the multiple-input single-output (MISO) case and in
\cite{LLPS-ITSubm09} for general MIMO case. Rather surprisingly, it
was shown in \cite{LLPS-ITSubm09} that, under the matrix power
constraint \eqref{eq:MC}, both confidential messages can be
\emph{simultaneously} communicated at their respected maximum
secrecy rates.
\item With only one confidential message ($W_1$ or $W_2$) \emph{and} the
common message $W_0$, the secrecy capacity region of the channel was
characterized in \cite{LLL-ITSubm09} using a channel-enhancement
approach \cite{WSS-IT06} and an extremal entropy inequality of
Weingarten {\it et al.} \cite{WLSSV-IT09}.
\end{itemize}
The main contribution of this paper is to provide a precise
characterization of the secrecy capacity region of the MIMO Gaussian
broadcast channel with a more complete message set that includes a
common message $W_0$ and two independent, confidential messages
$W_1$ and $W_2$ by generalizing the channel-enhancement argument of
\cite{LLL-ITSubm09}.

\section{Main Result} \label{sec:MAIN}

The main result of the paper is summarized in the following theorem.

\begin{theorem}[General MIMO Gaussian broadcast channel]\label{thm:GMBC}
The secrecy capacity region of the MIMO Gaussian broadcast channel
\eqref{eq:Ch1} with a common message $W_0$ (intended for both
receivers) and confidential messages $W_1$ (intended for receiver 1
but needing to be kept secret from receiver 2) and $W_2$ (intended
for receiver 2 but needing to be kept secret from receiver 1) under
the matrix power constraint \eqref{eq:MC} is given by the set of
nonnegative rate triples $(R_0,R_1,R_2)$ such that
\begin{align}
R_0&\le \min\left\{\frac{1}{2}\log
\left|\frac{\Hv_1\Sv\Hv_1^{\T}+\Iv_{r1}}{\Hv_1(\Sv-\Bv_0)\Hv_1^{\T}+\Iv_{r1}}\right|,\right.\notag\\
&~\qquad\left.\frac{1}{2}\log\left|\frac{\Hv_2\Sv\Hv_2^{\T}+\Iv_{r2}}{\Hv_2(\Sv-\Bv_0)\Hv_2^{\T}+\Iv_{r2}}\right|
\right\}\notag\\
R_1 &\le
\frac{1}{2}\log\left|\Iv_{r_1}+\Hv_1\Bv_1\Hv_1^{\T}\right|\notag\\
&~\qquad-\frac{1}{2}\log\left|\Iv_{r_2}+\Hv_2
\Bv_1\Hv_2^{\T}\right|\notag\\
\text{and} \qquad R_2 &\le
\frac{1}{2}\log\left|\frac{\Iv_{r_2}+\Hv_2(\Sv-\Bv_0)\Hv_2^{\T}}{\Iv_{r_2}+\Hv_2\Bv_1\Hv_2^{\T}}\right|\notag\\
&~\qquad-\frac{1}{2}\log\left|\frac{\Iv_{r_1}+\Hv_1(\Sv-\Bv_0)\Hv_1^{\T}}{\Iv_{r_1}+\Hv_1\Bv_1\Hv_1^{\T}}\right|
\label{eq:SCR}
\end{align}
for some $\Bv_0 \succeq 0$, $\Bv_1 \succeq 0$ and $\Bv_0+\Bv_1
\preceq \Sv$. Here, $\Iv_{r_k}$ denotes the identity matrix of size
$r_k \times r_k$ for $k=1,2$.
\end{theorem}

\begin{remark}\label{rmk:1}
Note that for any given $\Bv_0$, the upper bounds on $R_1$ and $R_2$
can be simultaneously maximized by a same $\Bv_1$. In fact, the
upper bounds on $R_1$ and $R_2$ are fully symmetric with respect to
$\Hv_1$ and $\Hv_2$, even though it is not immediately evident from
the expressions themselves.
\end{remark}

To prove Theorem~\ref{thm:GMBC}, we shall follow \cite{WSS-IT06} and
first consider the \emph{canonical} aligned case. In an
\emph{aligned} MIMO Gaussian broadcast channel \cite{WSS-IT06}, the
channel matrices $\Hv_1$ and $\Hv_2$ are square and invertible.
Multiplying both sides of \eqref{eq:Ch1} by $\Hv_k^{-1}$, the
channel can be equivalently written as
\begin{equation}
\Yv_k[m] =\Xv_k[m]+\Zv_k[m], \quad k=1,2 \label{eq:CH-A}
\end{equation}
where $\{\Zv_k[m]\}_m$ is an i.i.d. additive vector Gaussian noise
process with zero mean and covariance matrix
$\Nv_k=\Hv_k^{-1}\Hv_{k}^{-\T}$, $k=1,2.$ The secrecy capacity
region of the aligned MIMO Gaussian broadcast channel is summarized
in the following theorem.

\begin{theorem}[Aligned MIMO Gaussian broadcast channel]\label{thm:GMBC-A}
The secrecy capacity region $\Cc_s^{\rm [SBC]}(\Nv_1,\Nv_2,\Sv)$ of
the aligned MIMO Gaussian broadcast channel \eqref{eq:CH-A} with a
common message $W_0$ and confidential messages $W_1$ and $W_2$ under
the matrix power constraint \eqref{eq:MC} is given by the set of
nonnegative rate triples $(R_0,R_1,R_2)$ such that
\begin{align}
R_0&\le \min\left\{\frac{1}{2}\log
\left|\frac{\Sv+\Nv_1}{(\Sv-\Bv_0)+\Nv_1}\right|,\frac{1}{2}\log
\left|\frac{\Sv+\Nv_2}{(\Sv-\Bv_0)+\Nv_2}\right|
\right\}\notag\\
R_1 &\le
\frac{1}{2}\log\left|\frac{\Bv_1+\Nv_1}{\Nv_1}\right|-\frac{1}{2}\log\left|\frac{
\Bv_1+\Nv_2}{\Nv_2}\right|\notag\\
R_2 &\le\frac{1}{2}\log\left|\frac{
(\Sv-\Bv_0)+\Nv_2}{\Bv_1+\Nv_2}\right|-
\frac{1}{2}\log\left|\frac{(\Sv-\Bv_0)+\Nv_1}{\Bv_1+\Nv_1}\right|
\label{eq:SCR-A}
\end{align}
for some $\Bv_0 \succeq 0$, $\Bv_1 \succeq 0$ and $\Bv_0+\Bv_1
\preceq \Sv$.
\end{theorem}

Next, we prove Theorem~\ref{thm:GMBC-A} by generalizing the
channel-enhancement argument of \cite{LLL-ITSubm09}. Extension from
the aligned case \eqref{eq:SCR-A} to the general case \eqref{eq:SCR}
follows from the standard limiting argument \cite{WSS-IT06}; the
details are deferred to the extended version of this
work~\cite{LLPS-ITsumb10-BCCC}.

\section{Proof of Theorem~\ref{thm:GMBC-A}} \label{sec:pf}

\subsection{Achievability} \label{sec:AC}

The problem of a two-receiver discrete memoryless broadcast channel
with a common message and two confidential common messages was
studied in \cite{XCC-IT09}, where an achievable secrecy rate region
was given by the set of rate triples $(R_0,R_1,R_2)$ such that
\begin{align}
R_0 &\le \min[I(\Uv_0; \Yv_1),~I(\Uv_0, \Yv_2)]\notag\\
R_1 &\le
I(\Vv_1;\Yv_1|\Uv_0)-I(\Vv_1;\Vv_2,\Yv_2|\Uv_0)\notag\\
\text{and}\qquad R_2 &\le
I(\Vv_2;\Yv_2|\Uv_0)-I(\Vv_2;\Vv_1,\Yv_1|\Uv_0) \label{eq:ASRR-LP}
\end{align}
where $\Uv_0$, $\Vv_1$ and $\Vv_2$ are auxiliary random variables
such that
$(\Uv_0,\Vv_1,\Vv_2)\rightarrow\Xv\rightarrow(\Yv_1,\Yv_2)$ forms a
Markov chain. The scheme to achieve this secrecy rate region is a
natural combination of secret dirty-paper coding and superposition
coding. Thus, the achievability of the secrecy rate region
\eqref{eq:SCR-A} follows from that of (\ref{eq:ASRR-LP}) by setting
\begin{align*}
\Vv_1 &=\Uv_1+\Fv \Uv_2\\ \Vv_2 & =\Uv_2\\
\mbox{and} \quad \Xv &=\Uv_0+\Uv_1+\Uv_2
\end{align*}
where $\Uv_0$, $\Uv_1$ and $\Uv_2$ are three independent Gaussian
vectors with zero means and covariance matrices $\Bv_0$, $\Bv_1$ and
$\Sv-\Bv_0-\Bv_1$, respectively, and $$\Fv
:=\Bv\Hv_1^{\T}(\Iv_{r_1}+\Hv_1\Bv\Hv_1^{\T})^{-1}\Hv_1.$$ The
details of the proof are deferred to the extended version of this
work~\cite{LLPS-ITsumb10-BCCC}.

\subsection{The converse} \label{sec:Conv}
Next, we prove the converse part of Theorem~\ref{thm:GMBC-A}
assuming that $\Sv\succ0$. The case where $\Sv\succeq 0$, $|\Sv|=0$
can be found in the extended version of this
work~\cite{LLPS-ITsumb10-BCCC}.

Let
\begin{align}
f_0(\Bv_0)&=  \min\left\{\frac{1}{2}\log
\left|\frac{\Sv+\Nv_1}{(\Sv-\Bv_0)+\Nv_1}\right|,\right.\notag\\
&~\quad\left.\frac{1}{2}\log
\left|\frac{\Sv+\Nv_2}{(\Sv-\Bv_0)+\Nv_2}\right|
\right\}\notag\\
f_1(\Bv_1) &=
\frac{1}{2}\log\left|\frac{\Bv_1+\Nv_1}{\Nv_1}\right|-\frac{1}{2}\log\left|\frac{
\Bv_1+\Nv_2}{\Nv_2}\right|\notag\\
\text{and} \qquad f_2(\Bv_0,\Bv_1) &= \frac{1}{2}\log\left|\frac{
(\Sv-\Bv_0)+\Nv_2}{\Bv_1+\Nv_2}\right| \notag\\
&~\quad-
\frac{1}{2}\log\left|\frac{(\Sv-\Bv_0)+\Nv_1}{\Bv_1+\Nv_1}\right|.
 \label{eq:def-f1}
\end{align}
Then, the secrecy rate region \eqref{eq:SCR-A} can be rewritten as
\begin{align}
&\Rc_{in}:=\bigcup_{\Bv_0 \succeq 0, \Bv_1 \succeq 0, \Bv_0+\Bv_1
\preceq \Sv} \bigl\{(R_0,R_1,R_2)\bigl|\notag\\
&\begin{array}{l}
R_0\le f_0(\Bv_0),~R_1\le f_1(\Bv_1),~R_2\le f_2(\Bv_0,\Bv_1)\\
\end{array} \bigr\}. \label{eq:SCR-A2}
\end{align}
To show that $\Rc_{in}$ is indeed the secrecy capacity region of the
aligned MIMO Gaussian broadcast channel \eqref{eq:CH-A}, we will
consider proof by contradiction. Assume that
$(R_0^{\dag},R_1^{\dag},R_2^{\dag})$ is an achievable secrecy rate
triple that lies \emph{outside} the region $\Rc_{in}$. Since
$(R_0^{\dag},R_1^{\dag},R_2^{\dag})$ is achievable, we can bound
$R_0^{\dag}$ by
\begin{align*}
R_0^{\dag} &\le \min \left(
\frac{1}{2}\log\left|\frac{\Sv+\Nv_1}{\Nv_1}\right|,\;
\frac{1}{2}\log\left|\frac{\Sv+\Nv_2}{\Nv_2}\right|\right)= R_0^{\rm max}.
\end{align*}
Moreover, if $R_1^{\dag}=R_2^{\dag}=0$, then $R_0^{\rm max}$ can be
achieved by setting $\Bv_0=\Sv$ and $\Bv_1=0$ in \eqref{eq:SCR-A}.
Thus, we can find $\lambda_1\ge 0$ and $\lambda_2 \ge 0$ such that
\begin{align}
\lambda_1 R_1^{\dag}+\lambda_2 R_2^{\dag}=\lambda_1 R_1^{\star}+\lambda_2
R_2^{\star}+\rho
\end{align}
for some $\rho > 0$, where $\lambda_1 R_1^{\star}+\lambda_2 R_2^{\star}$ is
given by
\begin{align}
\max_{(\Bv_0,\; \Bv_1)}\qquad &\lambda_1 f_1(\Bv_1)+\lambda_2 f_2(\Bv_0,\Bv_1) \notag\\
\text{subject to} \qquad
f_0(\Bv_0)&\ge R_0^{\dag}\notag\\
\Bv_0 &\succeq 0\notag\\
\Bv_1 &\succeq 0\notag\\
\Bv_0+\Bv_1 &\preceq \Sv. \label{eq:op-A}
\end{align}
Let $(\Bv_0^{\star},\Bv_1^{\star})$ be an optimal solution to the
above optimization program (\ref{eq:op-A}). Then,
$(\Bv_0^{\star},\Bv_1^{\star})$ must satisfy the following
Karush-Kuhn-Tucker (KKT) conditions:
\begin{align}
&(\beta_1+\lambda_2)[(\Sv-\Bv_0^{\star})+\Nv_1]^{-1} +\beta_2[(\Sv-\Bv_0^{\star})+\Nv_2]^{-1}
  +\Mv_0\notag\\
&\qquad\qquad\qquad=\lambda_2[(\Sv-\Bv_0^{\star})+\Nv_2]^{-1}+\Mv_2 \label{eq:KKT-1}\\
&\quad(\lambda_1+\lambda_2)(\Bv_1^{\star}+\Nv_1)^{-1}+\Mv_1\notag\\
&\qquad\qquad\qquad=(\lambda_1+\lambda_2)(\Bv_1^{\star}+\Nv_2)^{-1}+\Mv_2
\label{eq:KKT-2}\\
&\Mv_0\Bv_0^{\star}=0, ~\Mv_1\Bv_1^{\star}=0,
~\text{and}~\Mv_2(\Sv-\Bv_0^{\star}-\Bv_1^{\star})=0 \label{eq:KKT-4}
\end{align}
where $\Mv_0$, $\Mv_1$ and $\Mv_2$ are positive semidefinite
matrices, and $\beta_k$, $k=1,2$, is a nonnegative real scalar such
that $\beta_k>0$ if and only if
\begin{align*}
\frac{1}{2}\log \left|\frac{\Sv+\Nv_k}{(\Sv-\Bv_0^{\star})+\Nv_k}\right| =
R_0^{\dag}.
\end{align*}
Hence, we have
\begin{align}
(\beta_1 &+\beta_2)R_0^{\dag}+\lambda_1 R_1^{\dag}+ \lambda_2
R_2^{\dag}\notag\\
&=\frac{\beta_1}{2}\log
\left|\frac{\Sv+\Nv_1}{(\Sv-\Bv_0^{\star})+\Nv_1}\right|+\frac{\beta_2}{2}\log
\left|\frac{\Sv+\Nv_2}{(\Sv-\Bv_0^{\star})+\Nv_2}\right|\notag\\
& \quad
+\lambda_1\left(\frac{1}{2}\log\left|\frac{\Bv_1^{\star}+\Nv_1}{\Nv_1}\right|
-\frac{1}{2}\log\left|\frac{ \Bv_1^{\star}+\Nv_2}{\Nv_2}\right|\right)\notag\\
&\quad+\lambda_2\left(\frac{1}{2}\log\left|\frac{(\Sv-\Bv_0^{\star})+\Nv_2}{\Bv_1^{\star}+\Nv_2}\right| \right.\notag\\
&\qquad\left.
-\frac{1}{2}\log\left|\frac{(\Sv-\Bv_0^{\star})+\Nv_1}{\Bv_1^{\star}+\Nv_1}\right|\right)+\rho.
\label{eq:WsumR-1}
\end{align}

Next, we shall find a contradiction to (\ref{eq:WsumR-1}) by
following the following three steps.

\subsubsection{Step~1--Split Each Receiver into Two Virtual Receivers}
Consider the following aligned MIMO Gaussian broadcast channel with four
receivers:
\begin{align}
\Yv_{1a}[m]& =\Xv[m]+\Zv_{1a}[m] \notag\\
\Yv_{1b}[m]& =\Xv[m]+\Zv_{1b}[m] \notag\\
\Yv_{2a}[m]& =\Xv[m]+\Zv_{2a}[m] \notag\\
\text{and}\qquad \Yv_{2b}[m]& =\Xv[m]+\Zv_{2b}[m] \label{eq:CH-A2}
\end{align}
where $\{\Zv_{1a}[m]\}$, $\{\Zv_{1b}[m]\}$, $\{\Zv_{2a}[m]\}$ and $\{\Zv_{2b}[m]\}$ are i.i.d. additive vector Gaussian noise processes with zero means and covariance matrices $\Nv_1$, $\Nv_1$, $\Nv_2$ and $\Nv_2$, respectively. Suppose that the transmitter has three independent messages $W_0$, $W_1$ and $W_2$, where $W_0$ is intended for both receivers $1b$ and $2b$, $W_1$ is intended for receiver $1a$ but needs to be kept secret from receiver $2b$, and $W_2$ is intended for receiver $2a$ but needs to be kept
secret from receiver $1b$. Note that the channel (\ref{eq:CH-A2}) has the same secrecy capacity region as the channel (\ref{eq:CH-A}) under the same power constraints.

\subsubsection{Step~2--Construct an Enhanced Channel} \label{sec:en}
Let $\Nt$ be a real symmetric matrix satisfying
\begin{align}
\Nt&:=\left(\Nv_1^{-1}+\frac{1}{\lambda_1+\lambda_2}\Mv_1\right)^{-1}.
\label{eq:def-Nt1}
\end{align}
Note that the above definition implies that
$\Nt \preceq \Nv_1.$ Since $\Mv_1\Bv_1^{\star}=0$, following \cite[Lemma~11]{WSS-IT06}, we have
\begin{align*}
(\lambda_1+\lambda_2)(\Bv_1^{\star}+\Nt)^{-1}&=(\lambda_1+\lambda_2)(\Bv_1^{\star}+\Nv_1)^{-1}+\Mv_1
\end{align*}
and
\begin{align}
|\Bv_1^{\star}+\Nt||{\Nv_1}|&=\left|{\Bv_1^{\star}+\Nv_1}\right||{\Nt}|.
\label{eq:Enh-3}
\end{align}
Thus, by (\ref{eq:KKT-2}), we obtain
\begin{align}
(\lambda_1+\lambda_2)(\Bv_1^{\star}+\Nt)^{-1}&=(\lambda_1+\lambda_2)(\Bv_1^{\star}+\Nv_2)^{-1}+\Mv_2.
\label{eq:Enh-4}
\end{align}
This implies that
$\Nt \preceq \Nv_2.$ 
Consider the following enhanced aligned MIMO Gaussian broadcast channel
\begin{align}
\Yt_{1a}[m]& =\Xv[m]+\Zt_{1a}[m] \notag\\
\Yv_{1b}[m]& =\Xv[m]+\Zv_{1b}[m] \notag\\
\Yt_{2a}[m]& =\Xv[m]+\Zt_{2a}[m] \notag\\
\text{and}\qquad \Yv_{2b}[m]& =\Xv[m]+\Zv_{2b}[m] \label{eq:CH-A3}
\end{align}
where $\{\Zt_{1a}[m]\}$, $\{\Zv_{1b}[m]\}$, $\{\Zt_{2a}[m]\}$ and
$\{\Zv_{2b}[m]\}$ are i.i.d. additive vector Gaussian noise processes with zero
means and covariance matrices $\Nt$, $\Nv_1$, $\Nt$ and $\Nv_2$, respectively.
Since $\Nt \preceq \{\Nv_1,\Nv_2\}$, we conclude that the secrecy capacity
region of the channel (\ref{eq:CH-A3}) is at
least as large as the secrecy capacity region of the channel (\ref{eq:CH-A2}) under the same power constraints.

Furthermore, based on (\ref{eq:Enh-4}), we have
\begin{align}
[(\Sv-\Bv_0^{\star}) & +\Nt](\Bv_1^{\star}+\Nt)^{-1}\notag\\
&=[(\Sv-\Bv_0^{\star})+\Nv_2](\Bv_1^{\star}+\Nv_2)^{-1} \label{eq:Enh-7}
\end{align}
and hence,
\begin{align}
\left|\frac{(\Sv-\Bv_0^{\star})+\Nt}{\Bv_1^{\star}+\Nt}\right|
&=\left|\frac{(\Sv-\Bv_0^{\star})+\Nv_2}{\Bv_1^{\star}+\Nv_2}\right|.
\label{eq:Enh-9}
\end{align}
Combining (\ref{eq:KKT-1}) and (\ref{eq:Enh-4}), we may obtain
\begin{align}
(\lambda_1 &+\lambda_2)[(\Sv-\Bv_0^{\star})+\Nt]^{-1}\notag\\
&= (\lambda_2+\beta_1)[(\Sv-\Bv_0^{\star})+\Nv_1]^{-1}\notag\\
&\quad +(\lambda_1+\beta_2)[(\Sv-\Bv_0^{\star})+\Nv_2]^{-1}
+\Mv_0. \label{eq:Enh-12}
\end{align}
Substituting (\ref{eq:Enh-3}) and (\ref{eq:Enh-9}) into (\ref{eq:WsumR-1}), we
have
\begin{align}
&(\beta_1+\beta_2)R_0^{\dag}+\lambda_1 R_1^{\dag}+ \lambda_2 R_2^{\dag} \notag\\
&=\frac{\beta_1}{2}\log
\left|\frac{\Sv+\Nv_1}{(\Sv-\Bv_0^{\star})+\Nv_1}\right|+\frac{\beta_2}{2}\log
\left|\frac{\Sv+\Nv_2}{(\Sv-\Bv_0^{\star})+\Nv_2}\right|\notag\\
&\quad
+\lambda_1\left(\frac{1}{2}\log\left|\frac{(\Sv-\Bv_0^{\star})+\Nt}{\Nt}\right|
-\frac{1}{2}\log\left|\frac{ (\Sv-\Bv_0^{\star})+\Nv_2}{\Nv_2}\right|\right)\notag\\
&\quad+\lambda_2\left(\frac{1}{2}\log\left|\frac{(\Sv-\Bv_0^{\star})+\Nt}{\Nt}\right|
-\frac{1}{2}\log\left|\frac{(\Sv-\Bv_0^{\star})+\Nv_1}{\Nv_1}\right|\right)\notag\\
&\quad+\rho.
\label{eq:WsumR-2}
\end{align}

\subsubsection{Step~3--Outer Bound for the Enhanced Channel}
In the following, we shall consider a four-receiver discrete
memoryless broadcast channel with a common message and two
confidential messages and provide a single-letter outer bound on the
secrecy capacity region.

\begin{theorem}[Discrete memoryless broadcast channel] \label{thm:DMC}
Consider a discrete memoryless broadcast channel with transition probability
$p(\yt_{1a}, \yv_{1b}, \yt_{2a},\yv_{2b}|\xv)$ and messages $W_0$ (intended for
both receivers $1b$ and $2b$), $W_1$ (intended for receiver $1a$ but needing to
be kept confidential from receiver $2b$) and $W_2$ (intended for receiver $2a$
but needing to be kept confidential from receiver $1b$). If both
\begin{align*}
\Xv \rightarrow \Yt_{1a} \rightarrow ( \Yv_{1b}, \Yv_{2b}) \quad \text{and}
\quad  \Xv \rightarrow \Yt_{2a} \rightarrow ( \Yv_{1b}, \Yv_{2b})
\end{align*}
form Markov chains in their respective order, then the secrecy capacity region
of this channel satisfies $\Cc_s^{\rm [DMC]} \subseteq \Rc_o$,
where $\Rc_o$ denotes the set of nonnegative rate triples $(R_0,R_1,R_2)$ such
that
\begin{align}
R_0 &\le \min[I(\Uv; \Yv_{1b}),~I(\Uv, \Yv_{2b})]\notag\\
R_1 &\le
I(\Xv;\Yt_{1a}|\Uv)-I(\Xv;\Yv_{2b}|\Uv)\notag\\
\text{and}\qquad R_2 &\le I(\Xv;\Yt_{2a}|\Uv)-I(\Xv; \Yv_{1b}|\Uv)
\end{align}
for some $p(\uv, \xv)=p(\uv)p(\xv|\uv)$.
\end{theorem}
The proof of Theorem~\ref{thm:DMC} can be found in the extended
version of this work \cite{LLPS-ITsumb10-BCCC}.

Now, we may combine Steps 1, 2 and 3 and consider an upper bound on
the weighted secrecy sum-capacity of the channel (\ref{eq:CH-A}). By
Theorem~\ref{thm:DMC}, for any achievable secrecy rate triple
$(R_0,R_1,R_2)$ for the channel (\ref{eq:CH-A}) we have
\begin{align}
(\beta_1 &+\beta_2)R_0+\lambda_1 R_1+ \lambda_2 R_2 \notag\\
& \le \frac{\beta_1}{2}\log \left|2\pi e (\Sv+\Nv_1) \right|+
\frac{\beta_2}{2}\log \left|2\pi e (\Sv+\Nv_2) \right|\notag\\
&\quad +\frac{\lambda_1}{2}\log
\left|\frac{\Nv_2}{\Nt}\right|+
\frac{\lambda_2}{2}\log\left|\frac{\Nv_1}{\Nt}\right|+\eta(\lambda_1,\lambda_2)
\label{eq:BSR-1}
\end{align}
where
\begin{align*}
\eta&(\lambda_1,\lambda_2):=\lambda_1 h(\Xv+\Zt_{1a}|\Uv)+ \lambda_2
h(\Xv+\Zt_{2a}|\Uv) \notag\\
&\quad- (\lambda_2+\beta_1)
h(\Xv+\Zv_{1b}|\Uv)-(\lambda_1+\beta_2) h(\Xv+\Zv_{2b}|\Uv).
\end{align*}
Note that $0 \prec \Nt \preceq \{\Nv_1,\Nv_2\}$, $0\prec
\Bv_0^{\star} \preceq \Sv$ and $\Bv_0^{\star}\Mv_0=0$. Using
\cite[Corollary~4]{WLSSV-IT09} and (\ref{eq:Enh-12}), we may obtain
\begin{align}
\eta(\lambda_1,\lambda_2) &\le (\lambda_1+\lambda_2)\log\left|2\pi e
(\Sv-\Bv_0^{\star})+\Nt\right|\notag\\
&\quad- (\lambda_2+\beta_1)\log\left|2\pi e
(\Sv-\Bv_0^{\star})+\Nv_1\right|\notag\\
&\quad - (\lambda_1+\beta_2)\log\left|2\pi e (\Sv-\Bv_0^{\star})+\Nv_2\right|.
\label{eq:BSR-2}
\end{align}
Combining  (\ref{eq:BSR-1}) and (\ref{eq:BSR-2}), for any achievable
secrecy rate triple $(R_0,R_1,R_2)$ for the channel (\ref{eq:CH-A})
we have
\begin{align}
(&\beta_1+\beta_2)R_0+\lambda_1 R_1+ \lambda_2 R_2 \notag\\
&\le \frac{\beta_1}{2}\log
\left|\frac{\Sv+\Nv_1}{(\Sv-\Bv_0^{\star})+\Nv_1}\right|+\frac{\beta_2}{2}\log
\left|\frac{\Sv+\Nv_2}{(\Sv-\Bv_0^{\star})+\Nv_2}\right|\notag\\
&\quad
+\lambda_1\left(\frac{1}{2}\log\left|\frac{(\Sv-\Bv_0^{\star})+\Nt}{\Nt}\right|
-\frac{1}{2}\log\left|\frac{ (\Sv-\Bv_0^{\star})+\Nv_2}{\Nv_2}\right|\right)\notag\\
&\quad+\lambda_2\left(\frac{1}{2}\log\left|\frac{(\Sv-\Bv_0^{\star})+\Nt}{\Nt}\right|
-\frac{1}{2}\log\left|\frac{(\Sv-\Bv_0^{\star})+\Nv_1}{\Nv_1}\right|\right)\notag\\
&<  (\beta_1+\beta_2)R_0^{\dag}+\lambda_1 R_1^{\dag}+ \lambda_2 R_2^{\dag}.
\end{align}
Clearly, this contradicts the assumption that the rate triple
$(R_0^{\dag},R_1^{\dag},R_2^{\dag})$ is achievable. Therefore, we
have proved the desired converse result for
Theorem~\ref{thm:GMBC-A}.

\begin{figure}[t]
 \centerline{\includegraphics[width=0.7\linewidth,draft=false]{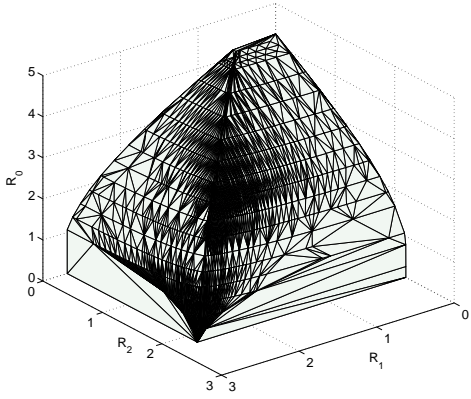}}
\caption{Secrecy capacity region $\{(R_0,R_1,R_2)\}$}
  \label{fig:sim1}
\end{figure}

\section{Numerical Example} \label{sec:ne}

In this section, we provide a numerical example to illustrate the
secrecy capacity region of the MIMO Gaussian broadcast channel with
a common message and two confidential messages. In this example, we
assume that both the transmitter and each of the receivers are
equipped with two antennas. The channel matrices and the matrix
power constraint are given by
\begin{align*}
\Hv_1 &=\left(\begin{matrix} 1.8&  2.0 \\ 1.0 &  3.0
\end{matrix}\right),~
\Hv_2 =\left(\begin{matrix} 3.3 &  1.3 \\
2.0 & -1.5
\end{matrix}\right)\\
\text{and} \qquad &\qquad
\Sv=\left(\begin{matrix} 5.0 & 1.25 \\ 1.25 & 10.0 \end{matrix}\right).
\end{align*}
which yield a \emph{nondegraded} MIMO Gaussian broadcast channel.
The boundary of the secrecy capacity region
$\Cc_s^{\rm[SBC]}(\Hv_1,\Hv_2,\Sv)$ is plotted in
Fig.~\ref{fig:sim1}.

\begin{figure}[t]
 \centerline{\includegraphics[width=0.7\linewidth,draft=false]{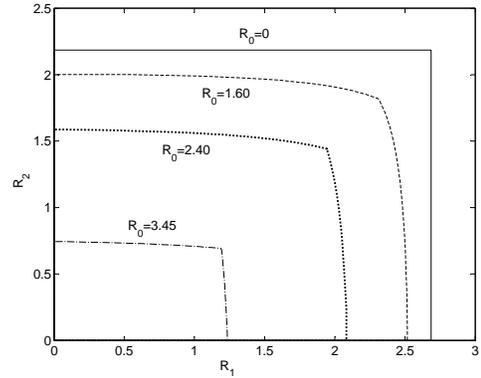}}
\caption{Secrecy rate regions $\{(R_1,R_2)\}$ for some given $R_0$}
  \label{fig:sim3}
\end{figure}

In Fig.~\ref{fig:sim3}, we have also plotted the boundaries of the
secrecy capacity region $(R_1,R_2)$ for some given common rate
$R_0$. It is particularly worth mentioning that with $R_0=0$, the
secrecy capacity region $\{(R_1,R_2)\}$ is \emph{rectangular}, which
implies that under the matrix power constraint, both confidential
messages $W_1$ and $W_2$ can be simultaneously transmitted at their
respective maximum secrecy rates. The readers are referred to
\cite{LLPS-ITSubm09} for further discussion of this phenomenon.

\section{Conclusion} \label{sec:CL}

In this paper, we have considered the problem of communicating two
confidential messages and a common message over a two-receiver MIMO
Gaussian broadcast channel. We have shown that a natural scheme that
combines secret dirty-paper coding and Gaussian superposition coding
achieves the entire secrecy capacity region. To prove the converse
result, we have applied a channel-enhancement argument and an
extremal entropy inequality of Weingarten {\it et al.}, which
generalizes the argument of \cite{LLL-ITSubm09} for the case with a
common message and only one confidential message.

\bibliographystyle{IEEEtran}
\bibliography{secrecy}

\end{document}